\documentclass[twocolumn,showpacs,showkeys,amsmath,amssymb,preprintnumbers,nofootinbib,floatfix,prb]{revtex4}

\usepackage{graphicx}
\usepackage{bm}
\usepackage{mathrsfs}
\usepackage{amsmath}
\usepackage{verbatim}
\usepackage[all, knot]{xy}
\xyoption{arc}

\begin{document}



\title{$k^{-3}$ superfluid spectrum of highly curved interacting quantum  vortices}

\author{Jeffrey Yepez
}


\address{Air Force Research Laboratory, Hanscom Air Force Base, Massachusetts  01731, USA
}

\begin{abstract}
Presented is a prediction, based on the Frenet-Serret differential geometry of space curves, that the wave number dependence of the average kinetic energy per unit length of two mutually interacting highly curved quantum vortex scales as $k^{-3}$.  The interacting quantum vortices are helical in shape, supporting circularly polarized counter-propagating waves, with arbitrary curvature and torsion.  This power-law spectrum agrees with  the high-$k$ spectrum found in precise quantum simulations of turbulent superfluidity with tangle of highly curved and excited quantum vortices.
\end{abstract}





\maketitle

Superfluid turbulence is an intriguing low-temperature phenomenon with power-law energy cascades that are undergoing active investigation.  In this Letter we consider the origin of a $k^{-3}$ power-law
 in the kinetic energy spectrum 
at high-$k$ wave numbers ($\gtrapprox$ the inverse coherence length) associated with highly curved quantum vortices.
The simplest theory for a superfluid condensate in the zero-temperature limit is
\begin{equation}
{\cal L}_\text{\tiny BEC}
 =
i \hbar\, \varphi^\ast
\partial_{t}
\varphi
+
 \frac{\hbar^2}{2m}
(\nabla  \varphi^\ast) \cdot
\nabla \varphi
+
\mu \,
\varphi^\ast 
 \varphi 
 -
\varphi^\ast V_\text{\tiny H}\,
 \varphi  ,
 \qquad
  \label{non_relativistic_Lagrangian_density}
\end{equation}
where $\varphi(x)$ is a complex scalar field for the Bose-Einstein condensate (BEC), $V_\text{\tiny H}$ is a local self-consistent Hartree potential, and
$\mu$ is the chemical potential.
Minimizing the action $\int d^4x\, {\cal L}_\text{\tiny BEC}$ leads to the Euler-Lagrange equation, a nonlinear Schroedinger equation
when $V_\text{\tiny H}(|\varphi|^2)=\frac{1}{2}g|\varphi|^2$, where $g$ is the real-valued coupling strength of the nonlinear interaction, known as the Gross-Pitaevskii equation (GPE)\cite{JMathPhys.1963.4.195,JETP.1961.2.451}
\(
\label{Gross_Pitaevskii_equation}
i \hbar\partial_t \varphi = - \frac{\hbar^2}{2m}\nabla^2 \varphi + (g | \varphi |^2 -\mu)\,\varphi.
\)
It captures complex vortex-vortex interplay (nucleation, emission and absorption of vortex rings, and reconnection) as well as quantum Kelvin wave mode (kelvon) production on the vortices and sound modes (phonons) that escape into the bulk region of the quantum fluid.

To determine a steady-state solution of the quantum vortex in a superfluid, with a BEC wave function denoted by $\varphi_\text{\tiny v}$, one solves the  time-independent GPE:
\(
\label{rescaled_GP_equation}
- \xi^2\nabla^2 \varphi_\text{\tiny v} + \left( \frac{g}{\mu}\, | \varphi_\text{\tiny v} |^2-1 \right)\varphi_\text{\tiny v}=0
\)
with
 healing length $\xi\equiv\hbar/\sqrt{2m \mu}$.
A solution for the background condensate wave function of a single rectilinear quantum vortex (with vorticity along $\hat{\bm{z}}$) is found by separation of variables in polar coordinates.  Inserting $\varphi_\text{\tiny v}(r,\vartheta,z)=\phi_\text{\tiny v}(r)Z_\text{\tiny v}(z) e^{i n\vartheta} $ into  time-independent GPE
with 
\(
g/\mu=\xi^2
\)
 gives the following equations with a separation constant $k_\parallel^2$:
\begin{subequations}
\begin{equation}
\label{parallel_part_of_GP_equation}
\frac{d^2 Z_\text{\tiny v}(z)}{dx^2} +k_\parallel^2 Z_\text{\tiny v}(z) =  0,
\vspace{-.10in}
\end{equation}
\vspace{-0.15in}
\begin{equation}
\label{radial_part_of_GP_equation}
\frac{d^2 \phi_\text{\tiny v}(r)}{dr^2} +\frac{1}{r}\frac{d \phi_\text{\tiny v}(r)}{dr}-\frac{n^2}{r^2}\phi_\text{\tiny v} (r) + \left(a-  \phi_\text{\tiny v}(r)^2\right) \phi_\text{\tiny v}(r) =  0,
\end{equation}
\end{subequations}
where $a\equiv \xi^{-2}-k_\parallel^2$.
Equation (\ref{parallel_part_of_GP_equation}) admits sinusoidal solutions and (\ref{radial_part_of_GP_equation}) can be solved for integer winding number $n$.  For the simplest $n=1$ case,  the Pad\'e approximant 
\(
\label{Pade_approximant}
\phi_\text{\tiny v} (r) =\sqrt{a} \sqrt{\frac{11 a r^2 (12 + a r^2)}{384+ a r^2 (128 + 11 a r^2)}}
\)
solves the
time-independent GPE
 with errors at ${\cal O}\left[(r\sqrt{a})^7\right]$.
Notice that $\phi_\text{\tiny v} (r)\rightarrow \sqrt{a}$ and $r\rightarrow \infty$, and thus the nonlinear term in (\ref{radial_part_of_GP_equation}) vanishes in the bulk.  

The rectilinear quantum vortex solution of 
the GPE
 is
\(
\varphi_\text{\tiny v}(x) = \phi_\text{\tiny v} (r) \left[Z_\circ^+ e^{ i (\omega_\parallel t+ k_\parallel z)}+Z_\circ^- e^{ i (\omega_\parallel t - k_\parallel z)}\right],
\)
with a parabolic dispersion relation $\hbar \omega_\parallel =\frac{\hbar^2 k_\parallel^2}{2m}$.
For a vortex line along the $\hat{\bm{z}}$-direction in cylindrical coordinates with unit winding number, the irrotational part of the  superfluid velocity has a divergent (perpendicular) part characteristic of inviscid flow and an advective  (parallel) part characteristic of rigid translation
\begin{eqnarray}
\label{v_field}
\bm{ v} 
\equiv
\frac{\hbar}{m}\nabla\left(\vartheta \pm  k_\parallel z\right)
\label{irrotational_part_of_superfluid_velocity_field_c}
=
 \frac{\hbar }{m r}\hat{\bm{ \vartheta}}
\pm
\frac{\hbar  k_\parallel}{m}\hat{\bm{z}}.
\end{eqnarray}
From this velocity field, we know that the circulation is quantized,
\(
\kappa \equiv \oint d\bm{l}\cdot \bm{v}
 =  \frac{h }{m}.
\)
From Stokes' theorem, we have
\(
\int d\bm{S} \cdot \nabla\times\bm{v} = \int d\bm{S} \cdot\bm{\omega}=  \frac{h }{m}.
\)
For a rectilinear  $\hat{\bm{z}}$-directed quantum vortex, the real part of the vorticity is pinned at the vortex center
\(
\label{vorticity_of_rectilinear_quantum_vortex}
\bm{\omega} =  \frac{h }{m} \delta^{(2)}(r) \hat{\bm{z}}.
\)
For a vortex filament of any shape, say a curve ${\cal C}$, the velocity field in general may be written as
\begin{equation}
\label{quantum_vortex_Biot_Savart_formula}
\bm{v}(\bm{r}) = \frac{\hbar}{2m}\oint_{\cal C} \frac{ d\bm{s}'\times (\bm{r}-\bm{s}')}{|\bm{r}-\bm{s}'|^3},
\end{equation}
where $d\bm{s}'$ is the differential length along the vortex filament, $\bm{s}'$ is the parametrization of ${\cal C}$, $\bm{r}$ is the field point, and  $d\bm{s}'$ is the differential line element at  the vortex center and parallel to the vorticity\cite{PhysRevB.31.5782}.   The Biot-Savart formula (\ref{quantum_vortex_Biot_Savart_formula}) reduces to  $\bm{v}=\hat{\bm{\vartheta}}\,{\hbar}/{(mr)}$ for the case of an infinite rectilinear quantum vortex positioned along the center of a cylindrical coordinate system with $\bm{s}'={\bm{z}}$, $d\bm{s}'=dz\, \hat{\bm{z}}$, and $d\bm{s}'\times(\hat{\bm{r}}-\hat{\bm{s}}')=\hat{\bm{\vartheta}} \,dz \, |z|/|\bm{r}-\bm{s}'|$.
    %


Using the Madelung transformation\cite{Madelung:1927} $\varphi = \sqrt{\rho}\,e^{iS/\hbar}$, (\ref{non_relativistic_Lagrangian_density}) can be written in terms of the conjugate fluid variables (the action $S$ and number density $\rho$) as follows:
\(
{\cal L}_\text{\tiny BEC} 
=
-\rho\, \partial_t S
-
 \rho
 \left[
 \frac{\left(\nabla S\right)^2}{2m}+ \frac{\hbar^2}{2m}\left(\frac{\nabla\rho}{2\rho}\right)^2 
-\mu
+V_\text{\tiny H}(\rho) 
\right] .
  \label{non_relativistic_Lagrangian_density_S_rho}
\)
The quantity in square brackets is identified with a Hamiltonian energy functional.
Bohm originally made this identification while considering quantum flow in a spatially-dependent linear external potential $V(\bm{x})$ \cite{PhysRev.85.166}.  For a BEC, the nonlinear internal potential energy is
\(
\label{internal_energy}
E_\text{\tiny int}(\rho) \equiv V_\text{\tiny H}(\rho)- \mu .
\)
A semiclassical  energy functional is thus
\(
\label{BEC_energy_functional}
H_\text{\tiny BEC} =  \frac{\left(\nabla S\right)^2}{2m}+ \frac{\hbar^2}{2m}\left(\frac{\nabla\rho}{2\rho}\right)^2 + E_\text{\tiny int}(\rho).
\)
%
The average energy is a statistical volume integral $\overline{E} = \int d^3 x \, 
 \rho\, H_\text{\tiny BEC} $ with $\rho$ taken as the effective probability distribution
\(
\label{conserved_average_energy}
\overline{E} 
=
  \int d^3 x \, 
 \rho
 \left[
 \frac{\left(\nabla S\right)^2}{2m}+ \frac{\hbar^2}{2m}\left(\frac{\nabla\rho}{2\rho}\right)^2 
 + E_\text{\tiny int}(\rho)\right].
\)
$\overline{E}$ is a conserved quantity.
Here we are interested in the first term
\( 
\overline{E}_\text{kin}^\text{cl}
= \int d^3 x \, 
 \rho
 \frac{\left(\nabla S\right)^2}{2m}
 =
 \int d^3 x
\,
\rho(x)\,
\left(\frac{1}{2}m
\bm{v}(x)^2
\right)
\),
where the de Broglie relation $m\bm{v} = \nabla S$ is used.
The average classical kinetic energy per unit length $L$ of a single linear quantum vortex is  ${\cal E} \equiv\overline{E}_\text{kin}^\text{cl}/L$:
\begin{equation}
\label{average_classical_kinetic_energy_per_unit_length}
{\cal E} 
=
m\rho_\circ  \int_{r_c}^{r_b} 2\pi r dr\, \frac{v_\vartheta^2}{2}
\stackrel{(\ref{irrotational_part_of_superfluid_velocity_field_c})}{=}
 \frac{\rho_\circ h^2}{4\pi m} \int_{r_c}^{r_b} \frac{dr}{r}
=
m\rho_\circ \frac{\kappa_\circ^2}{4\pi} \log \frac{r_b}{r_c},
\end{equation}
where $\kappa_\circ \equiv h/m$ is the quantum of circulation and $\rho_\circ$ is the constant background number density of the condensate, $r_b$ is a regularizing parameter associated with the size of the vessel containing a single quantum vortex
 and $r_c$ is an effective cutoff parameter to the divergent angular velocity field.\footnote{$r_c$ may be chosen so that the integral for ${\cal E}$, with a cutoff that avoids the singularity at the origin, is equivalent to the original nonsingular integral  with no cutoff.  So, $r_c$ is technically not a cutoff parameter per se because the original integral is nonsingular.  Instead, $r_c$ is merely a matching parameter useful for replacing a nonsingular but difficult integrand with an analytically simpler one.  All the expressions of the average classical kinetic energy per unit length in (\ref{average_classical_kinetic_energy_per_unit_length}) are equivalent and apply to a nearly straight (high-tension and low-curvature) vortex.}  


Consider the original treatment by Fetter for nearly-rectilinear vortices \cite{Fetter:1967p1942}.    The initial (single vortex) equilibrium states are parallel linear filaments with an arc length parametrization given by the following vectors
\(
\bm{R}_1^{(0)}=(\bm{r}_1, z_1)
\)
and 
\(
\bm{R}_2^{(0)}=(\bm{r}_2, z_2).
\)
The consequent deformed state due to the mutual interaction of the vortices is parametrized by
\begin{equation}
\label{deformed_R_s}
\bm{R}_1=(\bm{r}_1+\bm{u}_1(z_1), z_1)
\qquad
\qquad
\bm{R}_2=(\bm{r}_2+\bm{u}_2(z_2), z_2),
\end{equation}
where $\bm{u}_1$ and $\bm{u}_2$ are treated as small amplitude perturbations in the radial directions with respect to the initial unperturbed filamentary lines.  Each vortex filament is a stretched helix, approximating a nearly straight line parallel to the $\hat{\bm{z}}$-axis.

The first step is to calculate the fluctuation in position of a vortex element at the first vortex due to the presence of the second vortex.
The fluid velocity at the first vortex located at $\bm{R}_1$ caused by the second vortex at $\bm{R}_2$ is given by the Biot-Savart law (\ref{quantum_vortex_Biot_Savart_formula})
\begin{equation}
\label{2_vortex_Biot_Savart_formula}
\bm{v}(\bm{R}_1) = \frac{\kappa_\circ}{4\pi}\oint_{{\cal C}_2} \frac{ d\bm{s}_2\times (\bm{R}_1-\bm{R}_2)}{|\bm{R}_1-\bm{R}_2|^3}.
\end{equation}
With $i=1,2$ denoting the vortices, the differential arc length is
\(
\label{ds_change_rule}
d\bm{s}_i = \frac{d\bm{s}_i}{dz_i}dz_i = dz_i \left(\hat{\bm{z}}+ \frac{d\bm{u}_i(z_i)}{dz_i}\right).
\)
Then, according to (\ref{2_vortex_Biot_Savart_formula}), the fluctuation of the position of the vortex element originally at $\bm{R}^{(0)}_1$ is
\begin{subequations}
\label{velocity_fluctuation}
\begin{eqnarray}
\nonumber
\frac{d\bm{u}_1}{dt} 
\!\!\!
&\stackrel{(\ref{deformed_R_s})}{=}&
\!\!\!
 \frac{\kappa_\circ}{4\pi}\int dz_2
 \frac{ \left(\hat{\bm{z}} + \frac{d\bm{u}_2}{dz_2} \right) \times \left(\bm{r}_{12}+\hat{\bm{z}}z_{12} + \bm{u}_{12}\right)}
 {|\bm{r}_{12}+\hat{\bm{z}}z_{12} + \bm{u}_{12}|^3},
 \\
\label{velocity_fluctuation_c}
\end{eqnarray}
where $\bm{r}_{12}\equiv \bm{r}_{1}-\bm{r}_{2}$,  $z_{12}\equiv z_{1}-z_{2}$, and  $\bm{u}_{12}\equiv \bm{u}_{1}-\bm{u}_{2}$.
Also, defining $\bm{R}_{12}^{(0)}\equiv\bm{r}_{12}+\hat{\bm{z}}z_{12}$ and making use of the Taylor expansion
\(
\frac{1}{|\bm{R}_{12}^{(0)}+ \bm{u}_{12}|^3}
=
\frac{1}{|\bm{R}_{12}^{(0)}|^3}
-
\frac{3\bm{u}_{12}\cdot \bm{R}_{12}^{(0)}}{|\bm{R}_{12}^{(0)}|^5}
+
\cdots,
\)
we can write a leading order expansion of the position fluctuation (velocity of the first vortex element at $z_1$ due to the presence of the second vortex)
\begin{eqnarray}
\nonumber
\frac{d\bm{u}_1}{dt} 
\!\!\!
&\stackrel{(\ref{velocity_fluctuation_c})}{=}&
\!\!\!
 \frac{\kappa_\circ}{4\pi}\hat{\bm{z}}\times\int dz_2
 \Big[
 \frac{\bm{r}_{12}+\bm{u}_{12} -   z_{12}\frac{d\bm{u}_2}{dz_2}}
 {|\bm{R}_{12}^{(0)}|^3}
 \\
 \label{velocity_fluctuation_d}
 &&
 \hspace{0.75in}
 -
  \frac{3 \,\bm{r}_{12} \cdot\bm{u}_{12} }
 {|\bm{R}_{12}^{(0)}|^5}
\bm{r}_{12}
  +\cdots
  \Big].
\end{eqnarray}
\end{subequations}

Making the analogy to the mutual inductance of two line currents, the Neumann formula can be used to calculate the interaction energy of two vortices
\begin{equation}
\label{Neumann_mutual_interaction_formula}
E_{12} = 
m \rho_\circ\int_{r_c}d^3x \,\frac{\bm{v}(x)^2}{2}
\stackrel{(\ref{quantum_vortex_Biot_Savart_formula})}{=}
\frac{m\rho \kappa_\circ^2}{4\pi}
\oint_{{\cal C}_1}\oint_{{\cal C}_2}\frac{d\bm{s}_1\cdot d\bm{s}_2}{|\bm{R}_1-\bm{R}_2|}.
\end{equation}
The next step is to develop an expansion for the interaction energy
\(
E_{12} 
=
\frac{m\rho \kappa_\circ^2}{4\pi}
\int\int dz_1 dz_2
 \frac{ \left(1 + \frac{d\bm{u}_1}{dz_1}\cdot \frac{d\bm{u}_2}{dz_2} + \cdots \right)}
 {|\bm{R}_{12}^{(0)}+\bm{u}_{12}|},
 \qquad
\)
where the cross-terms vanish because, in the reference frame at the original center of the unperturbed $i$th vortex line along $\hat{\bm{z}}$, the motion of the perturbed filament is in the polar direction $\frac{d\bm{u}_i}{dz_i}/| \frac{d\bm{u}_1}{dz_1}| \approx \pm\hat{\bm{\vartheta}}$.
Employing Taylor's theorem, the denominator is expanded to second order
\(
 \frac{1}
 {|\bm{R}_{12}^{(0)}+\bm{u}_{12}|}
 =
 \frac{1}
 {|\bm{R}_{12}^{(0)}|}
 -
 \frac{\bm{r}_{12}\cdot \bm{u}_{12}}
 {|\bm{R}_{12}^{(0)}|^3}
 +
  \frac{3(\bm{r}_{12}\cdot \bm{u}_{12})^2}
 {2|\bm{R}_{12}^{(0)}|^5}
 -
 \frac{(\bm{u}_{12})^2}
 {2|\bm{R}_{12}^{(0)}|^3}
 +\cdots,
 \)
so in turn the interaction energy expansion becomes
\begin{eqnarray}
\nonumber
E_{12} 
\!\!\!
&=&
\!\!\!
\frac{m\rho \kappa_\circ^2}{4\pi}
\int\int dz_1 dz_2
\Big[
 \frac{1}
 {|\bm{R}_{12}^{(0)}|}
 -
 \frac{\bm{r}_{12}\cdot \bm{u}_{12} -
\frac{1}{2}(\bm{u}_{12})^2}
 {|\bm{R}_{12}^{(0)}|^3}
 \\
 \label{2_quantum_vortex_mutual_energy_expansion_d}
 &&
 +
 \frac{z_{12}}
 {|\bm{R}_{12}^{(0)}|^3}
 \frac{d\bm{u}_2}{dz_2} \cdot  \bm{u}_1
 +
  \frac{\frac{3}{2}(\bm{r}_{12}\cdot \bm{u}_{12})^2}
 {|\bm{R}_{12}^{(0)}|^5}
 +\cdots 
 \Big],
\end{eqnarray}
where in (\ref{2_quantum_vortex_mutual_energy_expansion_d}) the term obtained by integrated by parts was rewritten as $-\frac{d}{dz_1}\frac{1}{|\bm{R}_{12}^{(0)}|} = \frac{\hat{\bm{z}}\cdot\bm{R}_{12}^{(0)}}{|\bm{R}_{12}^{(0)}|^3}=\frac{z_{12}}{|\bm{R}_{12}^{(0)}|^3}$.  Hence, it is straightforward to calculate the variation of the mutual interaction energy with respect to a fluctuation at $z_1$ of the center of the first vortex line
\begin{equation}
\begin{split}
\frac{\delta E_{12} 
}{\delta \bm{u}_1(z_1)}
=&
 \frac{m\rho\kappa_\circ^2}{4\pi}\int dz_2
 \Big[
- \frac{\bm{r}_{12}+\bm{u}_{12} -   z_{12}\frac{d\bm{u}_2}{dz_2}}
 {|\bm{R}_{12}^{(0)}|^3}
 \\
 &+
  \frac{3 \,\bm{r}_{12} \cdot\bm{u}_{12} }
 {|\bm{R}_{12}^{(0)}|^5}
\bm{r}_{12}
  +\cdots
  \Big].
  \end{split}
  \end{equation}
Comparing this result with the previous result (\ref{velocity_fluctuation_d}) yields the useful relation
\begin{equation}
\label{local_frame_relation}
m\rho\kappa_\circ\frac{d\bm{u}_1(z_1)}{dt} 
=
-\hat{\bm{z}}\times \frac{\delta E_{12} 
}{\delta \bm{u}_1(z_1)}.
\end{equation}
The mutual interaction energy part of the condensate energy arising from a perturbed quantum vortex of length $L=\oint_{\cal C}\sqrt{dz^2 + d\bm{u}^2}$ is primarily due to its bending, assuming a sufficient separation distance exists between the vortices so that vortex-vortex straining has no low-order effect.  Therefore, we have
\begin{subequations}
\begin{eqnarray}
\frac{\delta E_{12} 
}{\delta \bm{u}_1(z_1)}
 & \stackrel{(\ref{Neumann_mutual_interaction_formula})}{\stackrel{(\ref{average_classical_kinetic_energy_per_unit_length})}{=}} & 
 {\cal E} \frac{\delta L 
}{\delta \bm{u}_1(z_1)}
 =  
- {\cal E} 
\,
\frac{d^2\bm{u}_1(z_1)}{dz_1^2}+\cdots
\end{eqnarray}
\end{subequations}
Inserting this result into (\ref{local_frame_relation}), yields
\begin{equation}
\label{local_frame_relation_2}
m\rho\kappa_\circ\frac{d\bm{u}_1(z_1)}{dt} 
=
{\cal E} \hat{\bm{z}}\times \frac{d^2\bm{u}_1(z_1)}{dz_1^2}.
\end{equation}
Having completed our review of Fetter's treatment, let us
consider a high-curvature generalization of (\ref{local_frame_relation_2}).


The Frenet-Serret formulas of multivariable calculus concerning the geometry of curves describe the kinematic properties of a particle at position $\bm{R}$ moving along a continuous and differentiable curve  $\cal C$ (the particle's trajectory or world line) embedded in three-dimensional Euclidean space $\mathbb{R}^3$ 
\label{Frenet_Serret_formulas}
\begin{eqnarray}
\label{Frenet_Serret_formulas_a}
\frac{d\hat{\bm{t}}}{ds} & = & \kappa \hat{\bm{n}} 
\qquad
\quad
\label{Frenet_Serret_formulas_b}
\frac{d\hat{\bm{n}}}{ds}  =  -\kappa \hat{\bm{t}} +\tau  \hat{\bm{b}}
\qquad
\quad
\label{Frenet_Serret_formulas_c}
\frac{d\hat{\bm{b}}}{ds}  =  -\tau \hat{\bm{n}} ,
\qquad
\end{eqnarray}
where $s$ is the arc length parameter along $\cal C$, 
$\hat{\bm{t}}$ is the unit tangent to $\cal C$ at the point $\bm{R}$, 
$\hat{\bm{n}}$ is the unit normal perpendicular to $\hat{\bm{t}}$ at $\bm{R}$, 
$\hat{\bm{b}}$ is the unit bi-normal perpendicular to both $\hat{\bm{t}}$  and $\hat{\bm{n}}$, 
and
$\kappa$ is the curvature and $\tau$ is the torsion  of $\cal C$ at $\bm{R}$.  
Given a fixed curve $\cal C$, one constructs a local Frenet-Serret frame as follows
\label{Frenet_Serret_construction}
\begin{eqnarray}
\label{Frenet_Serret_construction_a}
\hat{\bm{t}} \equiv \frac{ \bm{R}'(s)}{| \bm{R}'(s)|}
\qquad
\label{Frenet_Serret_construction_b}
\hat{\bm{n}} \equiv \frac{ \hat{\bm{t}}'(s)}{| \hat{\bm{t}}'(s)|}
\qquad
\label{Frenet_Serret_construction_c}
\hat{\bm{b}} \equiv  \hat{\bm{t}}\times \hat{\bm{n}},
\end{eqnarray}
 where the prime indicates differentiation with respect to $s$.  So $\hat{\bm{n}}$ points along the direction of the derivative of $\hat{\bm{t}}$ with respect to the arc length parameter of the curve and equating (\ref{Frenet_Serret_formulas_a}) with (\ref{Frenet_Serret_construction_b}), the curvature is
 \(
\kappa = | \hat{\bm{t}}'(s)|.
 \)
  Therefore, the unit vectors $\hat{\bm{t}}$, $\hat{\bm{n}}$, and $\hat{\bm{b}}$ serve as an orthogonal coordinate system centered at $\bm{R}$, a local reference frame that moves with the particle. For example, specifying the points in $\mathbb{R}^3$ with the cylindrical coordinates $(r, \vartheta, z)$, if $\cal C$ is a helix with its axis along $r=0$
\begin{subequations}
\label{particle_trajectory}
  \begin{eqnarray}
\bm{R}(s) &=& 
\label{particle_trajectory_a}
\left( \mathfrak{r} \cos k z, \mathfrak{r} \sin k z , 0\middle) + \middle(0,0, \mathfrak{h} k z\right)
\\
\label{particle_trajectory_c}
&=&
\bm{u}(k z) +
\begin{cases}
0+ {\cal O}\left(\frac{\mathfrak{h}}{\mathfrak{r}}\right)     & \mathfrak{h}\ll \mathfrak{r}, \\
s\hat{\bm{z}}-{\cal O}\left(\frac{\mathfrak{r}^2}{\mathfrak{h}^2}\right)      & \mathfrak{r} \ll \mathfrak{h}
\end{cases}
\end{eqnarray}
\end{subequations}
 with $k z\equiv s/\sqrt{\mathfrak{r}^2+\mathfrak{h}^2}$,  then the Frenet-Serret frame is
\begin{eqnarray}
\label{Frenet_Serret_frame}
\hat{\bm{t}} & = & \frac{\mathfrak{r} \hat{\bm{\vartheta}} + \mathfrak{h} \hat{\bm{z}} }{\sqrt{\mathfrak{r}^2+ \mathfrak{h}^2}}
\qquad
\hat{\bm{n}}  =  -\hat{\bm{r}}
\qquad
\hat{\bm{b}}  =  \frac{-\mathfrak{h} \hat{\bm{\vartheta}} + \mathfrak{r} \hat{\bm{z}} }{\sqrt{\mathfrak{r}^2+ \mathfrak{h}^2}}
\end{eqnarray}   
   %
and the local curvature and  torsion of ${\cal C}$ are 
\begin{eqnarray}
\label{curvature_and_torsion_of_helix}
\kappa = \frac{\mathfrak{r}}{\mathfrak{r}^2+ \mathfrak{h}^2}
\qquad
\qquad
\tau = \frac{ \mathfrak{h}}{\mathfrak{r}^2+ \mathfrak{h}^2}.
\end{eqnarray}
In the limit $ \mathfrak{h}\ll \mathfrak{r}$, the curve is a compressed helix where each cycle approximates a circle with curvature $1/\mathfrak{r}$.  In the opposite limit $ \mathfrak{h}\gg \mathfrak{r}$, the curve is a stretched helix (approximating a straight line) with torsion $1/ \mathfrak{h}$.

We may reconsider the case of a perturbation to a rectilinear quantum vortex:  a quantum vortex supporting a small amplitude circularly-polarized plane Kelvin wave counterrotating in a sense opposite to the vorticity direction of the unperturbed vortex line.   The solution we seek is based on the general Frenet-Serret formulas.\footnote{
To make additional contact with the previous literature,  consider the limit of small curvature with the perturbed vortex is a stretched helix.   The local frame,
   which we may choose to fix at  $z_1$ centered on the first quantum vortex, is
\(
\hat{\bm{t}} \equiv \frac{ \bm{u}_1'(z_1)}{| \bm{u}_1'(z_1)|} =\hat{\bm{z}}+\cdots
\), 
\(
\hat{\bm{n}} \equiv \frac{\bm{u}_1''(z_1)}{| \bm{u}_1''(z_1)|}
=-\hat{\bm{r}}+\cdots,
\) and
\(
\hat{\bm{b}} \equiv  \hat{\bm{t}}\times \hat{\bm{n}}
=-\hat{\bm{\vartheta}}+\cdots
\), 
where the arc length is parametrized by $s\approx z_1$.  
The mutual interaction of the vortices causes them to bend into the filamentary shape of a rotating helix---a circularly polarized Kelvin wave.  Equation (\ref{local_frame_relation_2}) may be rewritten as
\begin{equation*}
\label{Schwarz_local_induction_approximation}
\dot{\bm{u}} 
\stackrel{(\ref{Frenet_Serret_construction_a})}{=}
 \frac{{\cal E}}{m\rho \kappa_\circ}\frac{ \bm{u}'}{|\bm{u}'|} \times \bm{u}'',
\end{equation*}
 In a parametrization with $|\bm{u}'|=1$, 
 this is the local induction approximation (LIA) used for quantum turbulence simulations 
 \cite{PhysRevB.31.5782}.
}
Since the second derivative of the radial perturbation of the quantum vortex center is
\(
\bm{u}_1'' \stackrel{(\ref{particle_trajectory})}{=} -k^2 \bm{u}_1,
\)
 (\ref{local_frame_relation_2}) takes the form of an undamped Bloch equation
 \begin{equation}
\label{quasi_Block_equation}
\frac{d\bm{u}_1(z_1)}{dt} 
=
 \frac{ {\cal E} k^2}{m\rho\kappa_\circ}  \,\bm{u}_1(z_1) \times \hat{\bm{s}}_2.
\end{equation}
The radial displacement $\bm{u}_1$ behaves like the magnetization vector of a nuclear spin precessing about a background magnetic field along $\hat{\bm{s}}_2\approx\hat{z}$.  Thus, two segments of the mutually interacting quantum vortices behave like coupled nuclear spins.

The helix rotates in time as a  sinusoidal perturbation $\bm{u}_1 \sim \mathfrak{r}_1 \,e^{i(k z -\omega t)}$,
 so (\ref{local_frame_relation_2}) may be written as
\begin{eqnarray}
i \omega\bm{u}_1(z_1)
 & = &  
 \frac{ {\cal E}(k) k^2}{m\rho\kappa_\circ} \hat{\bm{z}}\times \bm{u}_1(z_1).
\end{eqnarray}
Then with $\bm{u}_1 = (x,y)$, we have
\(
i \omega (x,y)
 = 
 \frac{ {\cal E}(k) k^2}{m\rho\kappa_\circ}(-y,x),
\)
or in matrix form
\(
\scriptsize
\begin{pmatrix}
 i \omega     &   \frac{ {\cal E} k^2}{m\rho\kappa_\circ}  \\
  - \frac{ {\cal E} k^2}{m\rho\kappa_\circ}    &  i \omega
\end{pmatrix}
\begin{pmatrix}
      x    \\
     y  
\end{pmatrix}
=
\begin{pmatrix}
      0    \\
      0  
\end{pmatrix}.
\)
The well known solution by setting the determinant to zero, for $r_b = 1/k$, is the Kelvin wave dispersion relation 
\begin{eqnarray}
\omega_\text{\tiny K} & =  &  \frac{ {\cal E}(k) k^2}{m\rho\kappa_\circ}
\label{semiclassical_Kelvin_wave_dispersion_relation}
\stackrel{(\ref{average_classical_kinetic_energy_per_unit_length})}{=}
 \frac{\kappa_\circ k^2}{4\pi}  \log \frac{1}{r_c k},
\end{eqnarray}
valid in the limit where $\mathfrak{r}/\mathfrak{h}$ is a small quantity with
${\cal C}$ nearly a line.
Now, with a locally helical-shaped quantum vortex parametrized by (\ref{particle_trajectory}),  
we need not restrict  $\mathfrak{r}/\mathfrak{h}$  to be a small parameter.  Instead the only generic constraint
 imposed upon
  the shape of the perturbed quantum vortex is that it be locally parametrized by (\ref{particle_trajectory_a}).  
We will not loose the generality of two interacting quantum vortices of arbitrary shape by restricting their mutually interacting segments to helices.

Since $| \bm{R}'(s)|=1$, $\hat{\bm{t}} \stackrel{(\ref{Frenet_Serret_construction_a})}{=}  \bm{R}'(s)$ and in turn (\ref{Frenet_Serret_formulas_a}) is
\begin{equation}
\label{curvature_equation}
\frac{d^2 \bm{R}(s)}{ds^2} \stackrel{(\ref{curvature_and_torsion_of_helix})}{=} \frac{\mathfrak{r}}{\mathfrak{r}^2+\mathfrak{h}^2}\hat{\bm{n}}.
\end{equation}
So, allowing for high curvature, (\ref{Frenet_Serret_construction_c}) can be written as
\begin{subequations}
\begin{eqnarray}
 \hat{\bm{b}}
& \stackrel{(\ref{curvature_equation})}{=} & \frac{\mathfrak{r}^2+\mathfrak{h}^2}{\mathfrak{r}} \hat{\bm{t}}\times\frac{d^2 \bm{R}(s)}{ds^2} 
\\
& \stackrel{(\ref{particle_trajectory_a})}{=} & \frac{1}{\mathfrak{r}}\, \hat{\bm{t}}\times\bm{R}(s)
\\
\label{b_in_helix_formuation_e}
-\mathfrak{h} \hat{\bm{\vartheta}} + \mathfrak{r} \hat{\bm{z}} 
& \stackrel{(\ref{Frenet_Serret_frame})}{=} & 
\left( \hat{\bm{\vartheta}} +\frac{ \mathfrak{h}}{\mathfrak{r}} \hat{\bm{z}} \right)\times\bm{R}(s).
\end{eqnarray}
\end{subequations}
We are now considering a pair of parallel vortex segments, arbitrary helices equal in magnitude and spin, circling around each other whereby each segment moves by the influence of the velocity field of the other. 
Since the velocity field of a vortex segment (with a circularly-polarized radial perturbation) is anti-parallel to the bi-normal unit vector of a helical curve ({\it i.e.}, $\bm{v} \propto -\hat{\bm{b}}$),   this velocity field may be generally expressed in the Frenet-Serret frame as
\begin{eqnarray}
\label{v_in_helix_formation_c}
\bm{v}
 &\stackrel{(\ref{Frenet_Serret_frame})}{=}& 
\omega_\circ(\mathfrak{h} \hat{\bm{ \vartheta}}
-
\mathfrak{r}\hat{\bm{z}}).
\end{eqnarray}
Next, let us determine 
 the radius $\mathfrak{r}$ and the stretching $\mathfrak{h}$ of the helix 
directly in terms of the available physical parameters describing the mutually interacting quantum vortices, such as the separation distance between the vortices $2R_{12}^{(0)}$.  Let us denote the orbital radius at $R_\circ \equiv R_{12}^{(0)}$.  Thus, the magnitude of the polar velocity component is 
\(
\label{tangential_orbital_velocity}
v_\vartheta = \frac{\hbar}{m (2 R_\circ)}.
\)
If $\bm{v}_\vartheta = \bm{\omega}_\circ \times \bm{R}_\circ$, where $\bm{\omega}_\circ =\omega_\circ\hat{\bm{z}}$ is the effective orbital ``cyclotron'' spin vector, then the orbital angular frequency of the vortex pair is
\(
\label{orbital_angular_frequency}
\omega_\circ = \frac{v_\vartheta}{R_\circ}=\frac{\hbar}{2m R_\circ^2}=\frac{\kappa_\circ}{4\pi R_\circ^2}.
\)
The quantum vortex solution (\ref{irrotational_part_of_superfluid_velocity_field_c}) with a plane-wave phonon mode counter-propagating to $\bm{\omega}_\circ$ along the $-\hat{\bm{z}}$ direction with wave number $k=k_\parallel$ is concomitant to the quantum vortex kelvon mode 
\begin{eqnarray}
\bm{v} 
&\stackrel{(\ref{v_field})}{=}  & 
 \frac{\hbar }{2 m R_\circ}\hat{\bm{ \vartheta}}
-
\frac{\hbar  k}{m}\hat{\bm{z}}
\label{v_in_helix_formation_b}
=
 \omega_\circ\left(
  R_\circ \hat{\bm{ \vartheta}}
-
2 R_\circ^2 k\hat{\bm{z}}
\right).
\end{eqnarray}
Equating (\ref{v_in_helix_formation_b}) to (\ref{v_in_helix_formation_c}),
 the helical parameters $\mathfrak{h}$ and $\mathfrak{r}$
are thus analytically determined to be
\(
\label{helix_radius_and_stretch_parameters}
\mathfrak{h} = R_\circ 
\)
and
\(
\mathfrak{r} = 2 R_\circ^2 k.
\)
In turn, the curvature and torsion of the helix are
\(
\kappa 
  \stackrel{(\ref{curvature_and_torsion_of_helix})}{=} 
\frac{2k_\parallel}{1+ 4 k^2 R_\circ^2 }
\)
and
\(
\tau
  \stackrel{(\ref{curvature_and_torsion_of_helix})}{=} 
 \frac{1}{R_\circ}\,
 \frac{1}{1+ 4 k^2 R_\circ^2 } .
\)
One observes that the maximum curvature $\kappa_\text{max} = k$ occurs when $k=1/(2R_\circ)$ and the maximum torsion $\tau_\text{max}= 1/R_\circ$ occurs when $k=0$.  Both the curvature and torsion vanish in the limit of infinitely separated straight-line vortices.
%
%
Inserting (\ref{v_in_helix_formation_c}) into (\ref{b_in_helix_formuation_e}), one finds a precise  self-consistent equation governing the dynamics of a quantum vortex with large-amplitude helical wave with arbitrary wave number $k$ in mutual interaction with another quantum vortex
\begin{subequations}
\begin{equation}
\label{Yepez_velocity_formula}
\dot{\bm{R} }
=
-\omega_\circ
\left( \hat{\bm{\vartheta}} +\frac{ \hat{\bm{z}}}{2R_\circ k} \right)\times\bm{R}(s).
\end{equation}
For analytical continuation to Fetter's treatment in the high-tension limit $\mathfrak{h}\gg \mathfrak{r}$, the fluctuation of (\ref{Yepez_velocity_formula}) obeys
\begin{equation}
\label{Yepez_velocity_formula_radial_part}
\dot{\bm{u}}
 \stackrel{(\ref{particle_trajectory_c})}{=} 
-
\frac{\omega_\circ}{2 R_\circ k} \hat{\bm{z}} \times\bm{u}.
\end{equation}
\end{subequations}
Equating (\ref{Yepez_velocity_formula_radial_part}) to the Bloch equation (\ref{quasi_Block_equation}) provides a way to determine the wave number dependence ${\cal E}={\cal E}(k)$:
\begin{equation}
\label{theoretical_Yepez_wave_spectrum}
{\cal E}(k) =  \frac{m\rho\kappa_\circ \omega_\circ}{2 R_\circ k^3} 
=
 \left(\frac{m\rho\kappa_\circ^2}{8\pi R_\circ^3}\right) \, k^{-3}.
\end{equation}
A 
helical wave triggered by a $k$-mode axial phonon counter-propagates along the segment undergoing mutual interaction.
Only the inital axial phonon dispersion relation
\(
\omega = \frac{\kappa_\circ}{4\pi} k^2
\)
is consonant with the 
dispersion relation for $\omega_\text{\tiny K}$ given by (\ref{semiclassical_Kelvin_wave_dispersion_relation}) for a semiclassical Kelvin wave. 

Equation (\ref{theoretical_Yepez_wave_spectrum}) is an
analytical prediction, based on the Frenet-Serret differential geometry of space curves, that the helical wave spectrum scales as $k^{-3}$ for highly curved quantum vortices.  
This power-law was found in precise quantum simulations of superfluid turbulence supporting highly curved vortices\cite{yepez:084501}.  A $k^{-3}$ spectrum also arises from a single rectilinear quantum vortex \cite{nore:2644}, and so it has been recently suggested that this straight vortex power-law underlies the high-$k$ part of a turbulent superfluid spectrum \cite{arXiv:0911.1749}. The analysis presented here suggests that (\ref{theoretical_Yepez_wave_spectrum}), not (\ref{average_classical_kinetic_energy_per_unit_length}) originally discovered by Lord Kelvin in 1880 nor the Fourier spectrum of a single rectilinear quantum vortex, may be responsible for the high-$k$ spectrum because of the effect of mutually interacting  highly-curved quantum vortices characteristic of superfluid turbulence.

\end{document}